# Pulsed ESR Measurement of Coherence Times in Si:P at Very Low Temperatures

W.D. Hutchison[a,b], L.K. Alexander[a,b], N. Suwuntanasarn[a,b] and G.N. Milford[a,c]

[a] *ARC Centre for Quantum Computer Technology.*
[b] *School of Physical, Environmental and Mathematical Sciences, The University of New South Wales, ADFA, Canberra, ACT 2600, Australia.*
[c] *School of Engineering and Information Technology, The University of New South Wales, ADFA, Canberra, ACT 2600, Australia.*

A purpose built millikelvin pulsed x-band ESR system is used to measure spin decoherence times of phosphorus donor spins in 99.92% isotopically pure 28 silicon. The isolated P spin $T_2$ is estimated at 260 (50) ms at 4.2 K and 330 (100) ms at 0.9 K.

## 1. Introduction

Phosphorus donor atoms ($^{31}$P) in silicon (Si:P) are expected to have very long (both nuclear and electron) spin relaxation times at low temperature. This combined with the obvious compatibility of silicon with existing device fabrication technology, makes this system of interest as a potential basis for quantum computing (QC) [1]. In Si:P, the dephasing of the donor electron spin represents the decoherence time of the device (single qubit decoherence). Pulsed electron spin resonance (ESR) offers a convenient and most effective way to study this dephasing. The original pulsed ESR studies of Si:P were conducted decades ago [2,3]. However, since the renewed interest in Si:P for QC applications, further work has been done. In particular, a projected isolated spin decoherence time ($T_2$) of 60 ms in an epilayer of isotopically enriched $^{28}$Si at 7 K was reported [4] more recently. There is considerable potential for this number to be improved (i.e. lengthened). In particular improvements in the $^{28}$Si purity are important since secondary hyperfine interactions between the donor spin and nuclear spin of I = ½ $^{29}$Si nuclei are a significant source of line broadening and spin decoherence. Also reduced phosphorus concentrations are desirable in the quest to establish $T_2$ in the single spin limit when using an ensemble measurement. There is an intrinsic decoherence caused by ensemble rotation of the refocusing pulse in an electron spin echo sequence and although this effect, also know as instantaneous diffusion (ID), may in principle be removed though the projection technique described below, there is a limit to the practical deconvolution of components with vastly different time scales. Finally lowering the temperature also has potential. Not only does the spin lattice relaxation ($T_1$), which provides the upper limit on $T_2$, get extremely long in Si:P at low temperature, but recently it has been suggested that decoherence based on pairwise interactions such as dipolar interactions can be suppressed at very low temperatures [5].

With an aim to extend such Si:P decoherence time measurements we have assembled a millikelvin X-band pulsed ESR system [6,7]. Here we report on some preliminary Si:P decoherence time measurements using a bulk 99.92% pure $^{28}$Si wafer with doping of 5 x 10$^{15}$ P cm$^{-3}$ from which we estimate isolated spin $T_2$'s for phosphorous donor electron spins in excess of 100 ms.

## 2. Experimental details

The dilution fridge based pulsed ESR system used in this work has been described elsewhere [6,7]. The sample used was a piece of bulk, phosphorus doped, isotopically enriched $^{28}$Si ($^{28}$Si:P) with the concentration ~5 x 10$^{15}$ P cm$^{-3}$ as determined by Hall bar measurements. Electron spin echo (ESE) pulse sequences were performed with the typical (π/2-τ-π-τ-echo) set at (16ns-τ-32ns-τ–echo). The resonant frequency used was approximately 9.4 GHz, and we tune the system and magnetic field to resonate the higher field (lower g factor) resonance satellite of the Si:P hyperfine split doublet since this line is clear of any extraneous surface charge trap resonance lines. With our set-up, ESE can be carried out at temperatures down to 60 mK. However, to sensibly follow the trends in the coherence time and allow the use of light to reset the spins between pulse sequences, ESE measurements were carried out at 4.2 K, 0.9 K and 0.2 K.

As mentioned above, the spin lattice relaxation time ($T_1$) for Si:P becomes very long at low temperatures. For P concentrations below ~10$^{16}$ cm$^{-3}$ the rate varies with a $T^7$ power down to 2 K and continues at $T^1$ below 2 K, with $T_1$ reaching >1000 s at 1.2 K [8]. This represents a major obstacle for signal averaging in ESE experiments since a delay of ~5 times $T_1$ should be applied between each pulse sequence. However other workers [8] have shown that $T_1$ could be shortened dramatically by the application of above band gap light (> ~1.0 µm). In this work we used 1 s wide light bursts (20 mW of 532 nm green laser light directed down a plastic light guide), followed by a 60 s wait time, between each pulse sequence. This choice of sequence was based on a comprehensive study of the effect of light performed in our earlier work [6,7]. Such a sequence generated a significant shortening of the relaxation time at temperatures down to 0.9 K, but no effect on the resulting shape of the echo decay curves as compared to waiting for much longer long times between pulse trains.

The Si:P echo decay results are fitted with using the following expression:

$$V(\tau) = V_0 \exp[-(2\tau/T_M) - (2\tau/T_{SD})^n]$$

where $T_M$ is the ensemble exponential decay time constant, which incorporates several terms detailed below, $T_{SD}$ is the spectral diffusion (SD) decay time associated mainly with interaction with $^{29}$Si nuclei and $n$ is an exponential stretching factor varied between 2 and 3 for different SD regimes by different authors [2,9]. The intrinsic phase memory time, $T_2$, also defined as a decoherence time of an isolated electron spin free from the effect of ID, is derived from $T_M$ and $T_{ID}$ via the relationship $1/T_M = 1/T_2 + 1/T_{ID}$. Where $T_{ID}$ may be estimated from $1/T_{ID} = C\pi\mu g^2\mu_B^2\sin^2(\beta/2)/(9\sqrt{3}\hbar)$ where $C$ is the concentration of the excited electron spins (for the concentration P in the sample, [P] = 2$C$), $\mu$ is the permeability of crystalline silicon, $g$ is the $g$-factor of the donor electron, $\mu_B$ is the Bohr magneton and β is the turn angle of the refocusing pulse.

In the case of the highly isotopically pure sample used here, we find that the $T_{SD}$ term can be largely ignored based on the fact that the echo decays can be fitted well to obtain estimates of $T_M$ with a single exponential function. To estimate $T_2$, we use the approach used by [4] in Si:P, and originally developed by [10] where it was recognised that since $T_{ID}$ is proportional sin$^2$(β/2) it is better to carry out a series of experiments with different values of β and then project to β = 0 to find the value of $T_2$ rather than relying on a multi-parameter fit of a single data set.

### 3. Results

Echo decay data collected for the bulk $^{28}$Si:P sample at temperatures of 4.2K, 0.9 K and 0.2 K with various second pulse turn angles are shown in Figs 1, 2 and 3 respectively. The resulting ensemble $T_M$ values are also listed on these figures. There is a trend apparent that the $T_M$ values for 0.9 K are longer than at 4.2 K for the same value of β. However the data at 0.2 K is much noisier and does not show clear trends. We merely include this later data here to demonstrate the capability of the system, but presume that the light resetting regime fails at this lower temperature. The 4.2 and 0.9 K data are, therefore only used to generate the $1/T_M$ v $\sin^2(\beta/2)$ plots of Fig 4. From such plots a linear projection to the $\sin^2(\beta/2) = 0$ axis yields estimates of the isolated $T_2$ values. From this data we obtain $T_2$ = 260 (50) ms at 4.2 K and 330 (100) ms at 0.9 K.

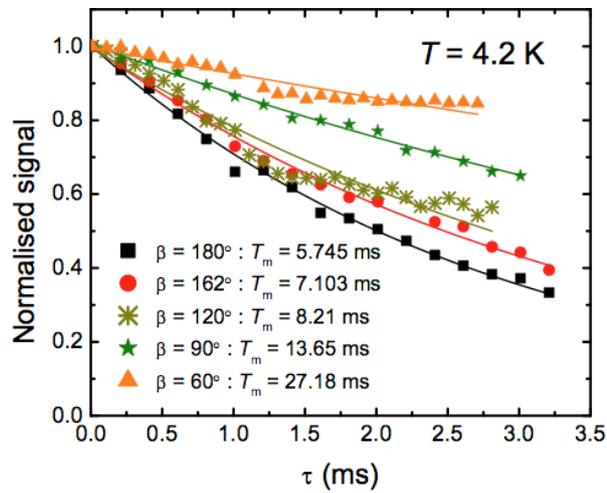

Fig. 1. Si:P ESE decay data for various refocusing pulse turn angles (β) at 4.2 K.

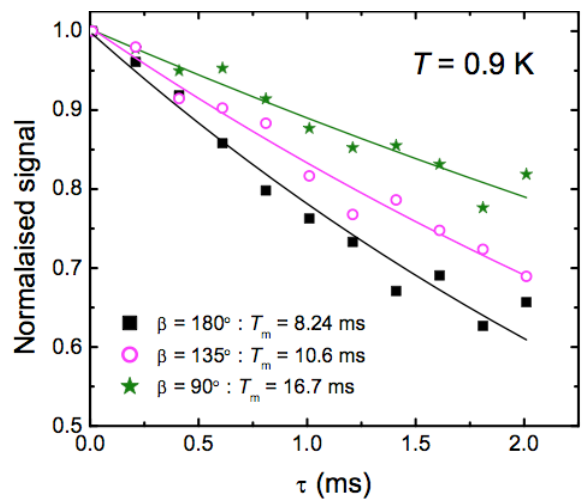

Fig. 2. Si:P ESE decay data for various refocusing pulse turn angles (β) at 0.9 K.

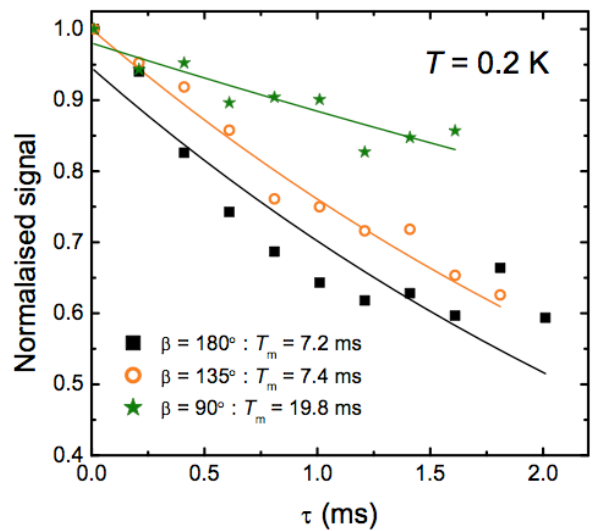

Fig. 3. Si:P ESE decay data for various refocusing pulse turn angles (β) at 0.2 K.

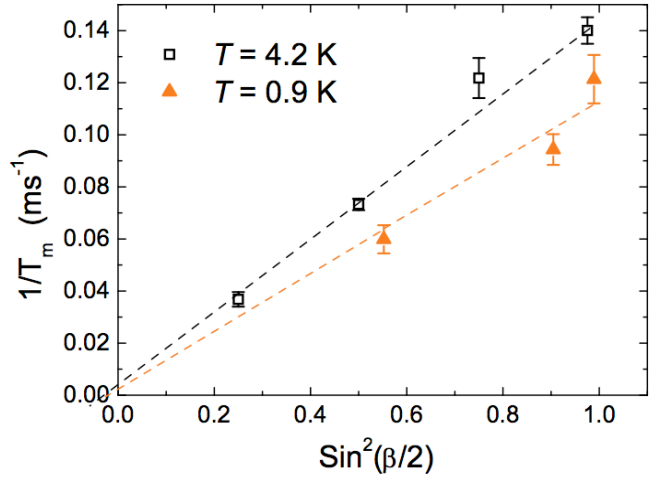

Fig. 4. $1/T_M$ v $\sin^2(\beta/2)$ plots from the ESE data at 4.2 and 0.9 K.

## 4. Discussion

In this paper we have presented ESE data collected for a bulk $^{28}$Si:P from which, using a projection to zero second pulse tip angle to remove the effect of instantaneous diffusion (ID), values of the isolated decoherence times of ~260 ms and ~330 ms are found at 4.2 K and 0.9 K respectively. Additional data at 0.2 K shows the capability of our system but indicates that the use of high power visible light to reset spins is problematic at low temperatures. Additionally it is noted that the theory of ID would suggest that the slope of $1/T_M$ v $\sin2(\beta/2)$ plots should depend only on the P concentration. It is curious therefore that the 4.2 K and 0.9 K data show apparently different slopes and that both of these correspond to values of C considerably lower than the $\sim 2 \times 10^{15}$ cm$^{-3}$ expected to match the Hall bar estimate for the sample. While it is common place for ESE estimated concentrations to be somewhat lower than that from bulk resistivity type measurements [6,7,9,11], in this case they seem excessively so at $4 \times 10^{14}$ and $3.5 \times 10^{14}$ cm$^{-3}$ respectively. We conclude that the low absolute values for reflect effective concentrations reduced by the light resetting process not accessing all spins. This process would also appear to have some residual temperature dependence. Better measurements may be possible if near band gap light (around 1 µm) is used instead of visible light as was found to be optimal at shortening $T_1$ in CWESR by [8].

**Acknowledgments**

The authors wish to thank Prof. M.S. Brandt for the loan of the sample. This work is supported by the Australian Research Council and by The University of New South Wales.